\documentclass[pra,twocolumn,showpacs,superscriptaddress,preprintnumbers,amsmath,amssymb,aps]{revtex4-1}
\usepackage{mathrsfs}
\usepackage{amsmath}
\usepackage{amssymb}
\usepackage{graphicx}
\usepackage{dcolumn}
\usepackage{bm}
\usepackage{subfigure}
\usepackage{color}
\usepackage{ifpdf}
\usepackage{epstopdf}
\usepackage{CJK}
\usepackage{verbatim}
\usepackage{bbm}
\usepackage{multirow}
\usepackage{booktabs}
 \usepackage{graphicx}
\usepackage{color}
\usepackage[
pdfstartview=FitH,%
bookmarks=true,%
bookmarksopen=true,%
breaklinks=true,%
colorlinks=true,%
linkcolor=blue,anchorcolor=blue,%
citecolor=blue,filecolor=blue,%
menucolor=blue,pagecolor=blue,%
urlcolor=blue]{hyperref}
\usepackage{float}
\usepackage[toc,page,title,titletoc,header]{appendix}
\usepackage{ragged2e}
\usepackage{lipsum}

\usepackage{array}
\usepackage{booktabs}

\usepackage{color}

\def\be{\begin{equation}}
	\def\ee{\end{equation}}
\def\bea{\begin{eqnarray}}
	\def\eea{\end{eqnarray}}

\begin{document}
		
\title{Collisional scattering of strongly interacting $D$-band Feshbach molecules in optical lattices}
\author{Fansu Wei}
\affiliation{State Key Laboratory of Advanced Optical Communication Systems and Networks, School of Electronics, Peking University, Beijing 100871, China}
\author{Chi-Kin Lai}
\affiliation{State Key Laboratory of Advanced Optical Communication Systems and Networks, School of Electronics, Peking University, Beijing 100871, China}
\author{Yuying Chen}
\affiliation {School of Physics and Electronics Engineering, Shanxi University, Taiyuan 030006, China}
\author{Zhengxi Zhang}
\affiliation{State Key Laboratory of Advanced Optical Communication Systems and Networks, School of Electronics, Peking University, Beijing 100871, China}
\author{Yun Liang}
\affiliation{State Key Laboratory of Advanced Optical Communication Systems and Networks, School of Electronics, Peking University, Beijing 100871, China}
\author{Hongmian Shui}
\affiliation{State Key Laboratory of Advanced Optical Communication Systems and Networks, School of Electronics, Peking University, Beijing 100871, China}
\affiliation{Institute of Carbon-based Thin Film Electronics, Peking University, Shanxi, Taiyuan 030012, China}
\author{Chen Li}\email{chen.li@tuwien.ac.at}
\affiliation{Vienna Center for Quantum Science and Technology, Atominstitut, TU Wien, Stadionallee 2, 1020 Vienna, Austria}
\author{Xiaoji Zhou}\email{xjzhou@pku.edu.cn}
\affiliation{State Key Laboratory of Advanced Optical Communication Systems and Networks, School of Electronics, Peking University, Beijing 100871, China}
\affiliation{Institute of Carbon-based Thin Film Electronics, Peking University, Shanxi, Taiyuan 030012, China}
\date{\today}
    \begin{abstract}
        The excited bands in optical lattices manifest an important tool for studying quantum simulation and many-body physics, making it crucial to measure high-band scattering dynamics under strong interactions. This work investigates both experimentally and theoretically the collisional scattering of $^{6}\rm Li_2$ molecular Bose-Einstein condensate in the $D$ band of a one-dimensional optical lattice, with interaction strength directly tunable via magnetic Feshbach resonance. We find a clear dependence of the $D$-band lifetimes on the interaction strength within the strongly interacting regime, which arises from the fact that the scattering cross-section is proportional to the square of the scattering length. The maximum lifetime versus lattice depth is measured to reveal the effects of interactions. We also investigate the scattering channels of $D$-band molecules under different interaction levels and develop a reliable two-body scattering rate equation. This work provides insight into the interplay between interaction and the collisional scattering of high-band bosons in optical lattices, paving the way for research into strong correlation effects in high-band lattice systems.
    \end{abstract}
 \maketitle

    \section{Introduction}
     Quantum gases confined in optical lattices have emerged as a versatile platform for exploring fundamental physics, particularly in many-body systems, due to their high controllability and robustness~\cite{RevModPhys.80.885}. These systems allow for the simulation of complex phenomena, such as phase transitions~\cite{Greiner2002,Jordens2008,Simon2011,Mazurenko2017}, quantum magnetism~\cite{doi:10.1126/science.1207239,doi:10.1126/science.1236362}, and strongly correlated systems~\cite{doi:10.1126/science.1165449,doi:10.1126/science.aaa7432,Tai2017}. An intriguing aspect of these systems is the study of excited bands, which provide insights into the production of novel quantum phases~\cite{Haldanefer, Li2014, PhysRevLett.99.200405, PhysRevLett.121.265301, PhysRevLett.126.035301,Wang2021} and the dynamics of external states~\cite{Kiefer2023, PhysRevLett.122.010402, PhysRevA.107.023303}.

    Collisional scattering is one of the most fundamental interaction processes in many-body systems. Research has focused on low-energy collisions within various systems, including atomic~\cite{Kemper_1984,PhysRevLett.81.938,RevModPhys.71.1,PhysRevLett.91.163201,PhysRevLett.103.163201,RN1701}, ionic~\cite{PhysRevLett.109.233202,PhysRevA.95.032709}, and electronic systems~\cite{SEITOV202026,RN1703,RN1704}. This process is particularly significant in the context of quantum gases, where the system's lifetime is strictly inversely proportional to the two-body collision rate. The scattering cross-section, which characterizes the two-body collision rate, has been extensively investigated both experimentally and theoretically in one-, two-, and mixed-dimensional optical lattices~\cite{PhysRevA.72.053604,RN1698,PhysRevA.88.033615,PhysRevLett.104.153202,PhysRevA.87.063638,PhysRevLett.111.205302}, such as the scattering model for atoms in the $P$ band of a 1D lattice~\cite{PhysRevLett.111.205302}, the measurement of the collision rate for atoms in the $D$ band of a 2D triangular lattice~\cite{PhysRevA.104.033326,Shui:23} and the observation of scattering halos~\cite{RN1697,Burdick_2016}, etc.

    In most cases, quantum simulation based on the excited bands~\cite{ramsey,PhysRevA.104.L060601,PhysRevA.109.043313,PhysRevResearch.6.023217} is significantly influenced by the collisional scattering regardless of the interaction levels. However, previous works have only studied the impact of parameters such as lattice depth and gas temperature on collision scattering in the weakly interacting regime. Currently, due to the difficulty of accurately tuning interactions in condensed excited-band systems in optical lattices, there is no reliable experimental evidence to verify the correspondence between interactions and excited band collision scattering rates within the strongly interacting regime.

    In this study, we explore both experimentally and theoretically the collisional scattering and lifetimes of $^{6}\mathrm{Li}_2$ molecular Bose-Einstein condensates (mBEC) in the $D$ band of a one-dimensional (1D) optical lattice. Our main focus is on the impact of inter-particle interactions on scattering rates and processes, with the interaction strength precisely adjustable via magnetic Feshbach resonance. We present measurements of the lifetimes of $D$-band molecules under various inter-particle interactions and lattice depths, revealing the squared relationship between $D$-band scattering rates and scattering lengths within the strongly interacting regime. The lattice depth with a maximum $D$-band lifetime is observed to shift with interaction strengths. Furthermore, we examine the $D$-band scattering processes in both strongly interacting $^{6}\mathrm{Li}_2$ system and weakly interacting $^{87}\mathrm{Rb}$ system, with the latter demonstrating strong agreement with the rate equation model developed in this study. We discuss the discrepancies between the two in depth and qualitatively analyze the interactions' role.

    This paper is organized as follows. In Sec.~\ref{sec:theoretic}, the theoretical model for the collisional scattering of $D$-band particles in a 1D optical lattice is described. Our experimental procedure, including the shortcut loading into the $D$ band and lifetime measurement, is shown in Sec.~\ref{sec:experimentImplementation}. In Secs.~\ref{sec:interaction}, we present the experimental results of $D$-band lifetimes with varying interaction strengths and lattice depths. The experimental results of scattering channels and the rate equation model are described in Sec.~\ref{sec:process} with a discussion. Finally, we give the conclusion in Sec.~\ref{sec:Conclusion}.

    \begin{figure}
    \includegraphics[width=0.48\textwidth]{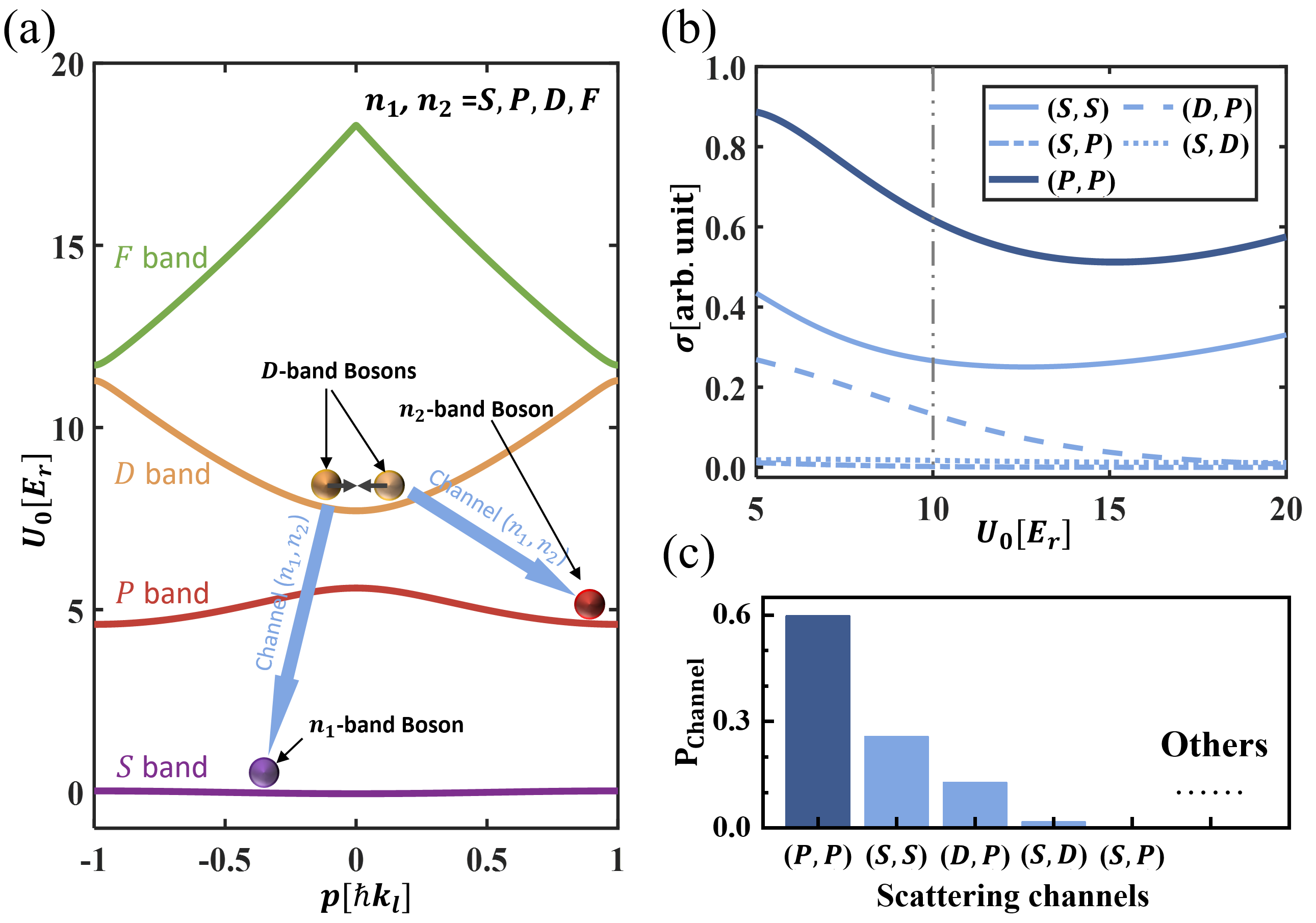}
    \caption{
		(a) Band structure of a 1D lattice and two-body scattering processes in the $D$ band. Two $D$-band bosons collide and each transitions into separate bands with lower total energy. (b) Relative scattering cross-sections are given by the theoretical model. The dark blue solid line marks the relative scattering cross-sections of $(D,D)$ to $(P,P)$, and the shallow blue solid line, dashed line, dotted line, and dash-dotted line mark $(D,D)$ to $(S,S)$, $(D,P)$, $(S,D)$ and $(S,P)$, respectively, under different lattice depths. (c) The relative proportion of channels with the lattice depth $U_0=10E_r$. The proportions of other channels are close to zero and can be considered negligible. Note that these scattering cross-sections are calculated in weak-interacting conditions.}
	\label{fig:Theo}
    \end{figure}
    \begin{figure*}
	\includegraphics[width=0.8\textwidth]{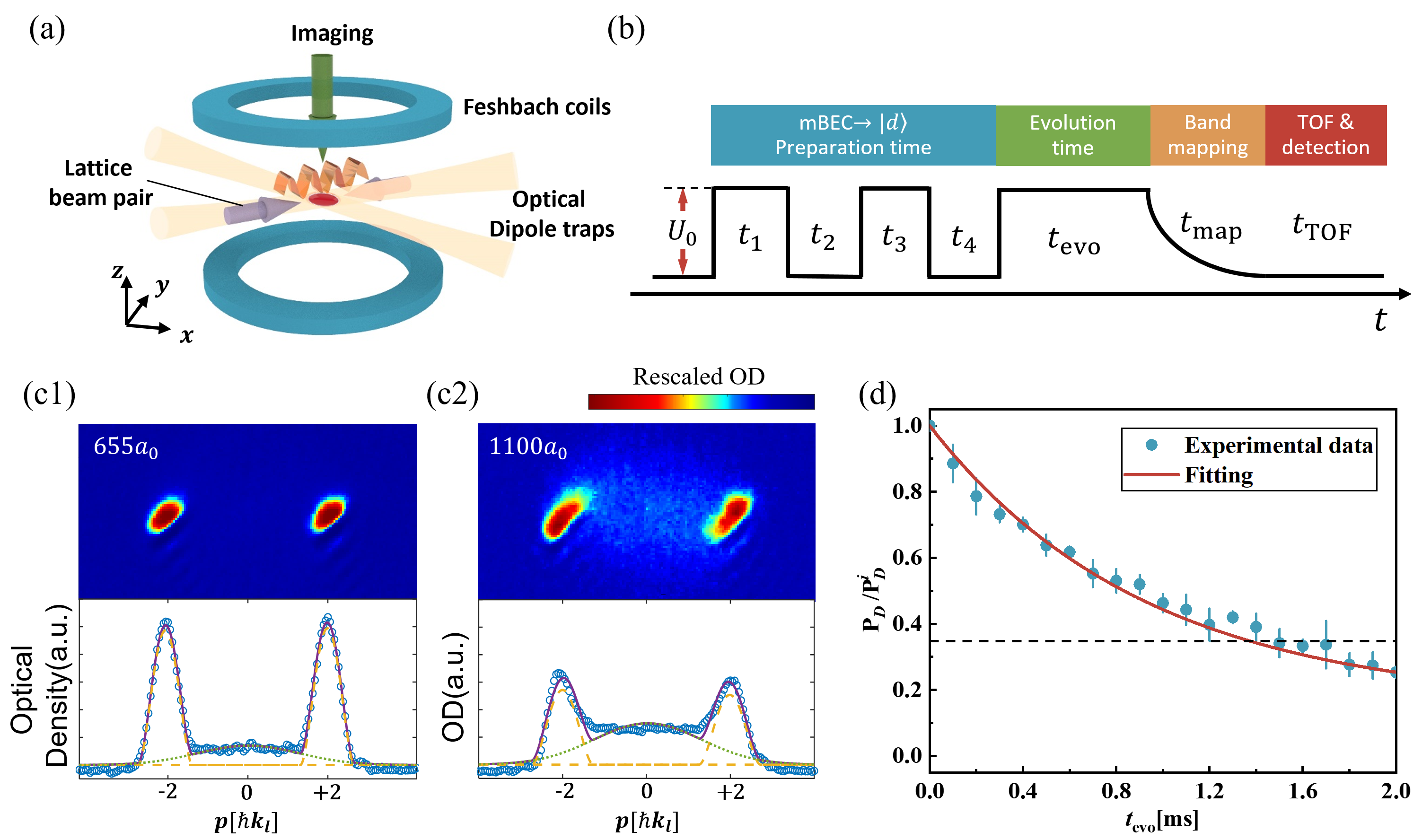}
	\caption{
		(a) Schematic of the experimental system. Feshbach mBECs are trapped in a pair of crossover dipole traps. The blue circles in the $x$-$y$ plane represent the Feshbach magnetic coils and the yellow Gaussian beams along the $x$-axis mark the crossed optical dipole trap. The two purple arrows in the $x$-$y$ plane stand for the counter-propagating lattice beams. The green arrow shows the imaging direction, which is perpendicular to the $x$-$y$ plane. (b) A typical experimental sequence of the lattice light. After preparing steady mBEC with varying interaction strengths, the lattice sequence will be initiated. The pulses applied before the evolution time $t_{\rm evo}$ are shortcut pulses designed to load mBEC into the $D$ band with lattice depth $U_0$. The lattice beam intensity decreases to zero adiabatically in time $t_{\rm map}$ to give a band mapping after $t_{\rm evo}$. $t_{\rm TOF}$ represents the time of flight following the complete switch-off of the lattice and trapping potential, after which imaging detection is performed. Experimental stages are shown above. (c) (top)Images taken with $U_0 = 10E_r$, $t_{\rm evo}=100\rm\ \mu s$, $t_{\rm map} = 100\rm\ \mu s$ and $t_{\rm TOF}=2\rm\ ms$ at different interaction strength and (bottom) the corresponding fitting result. (c1) $a_s=655a_0$. (c2) $a_s=1100a_0$. (d) Measurement of molecular lifetime in the $D$ band ($U_0 = 10E_r$, $a_s = 655a_0$). The $D$-band proportion $\mathrm{P}_D$ is normalized to $\mathrm{P}^i_D\simeq 0.7$, and the error bar shows the standard error of five measurements. The position of $\mathrm{P}_D/\mathrm{P}^i_D=1/e$ is marked by the gray dashed line.).
        }
	\label{fig:experimentalSetup}
    \end{figure*}
    
    \section{Theoretical model of excited-band scattering}\label{sec:theoretic}
    
    For ultracold bosons in the excited band of optical lattices, two-body collisions induced by $s$-wave scattering serve as the primary decay mechanism. In these collisions, two bosons in excited-energy bands collide and each scatters into the energy states with lower total energy. Here, we introduce the two-body scattering model in a 1D optical lattice, with the Hamiltonian given by
    \begin{align}
        \hat{H}=&-\frac{\hbar^2}{m}\left(\frac{\partial^2}{\partial y^2}+\frac{\partial^2}{\partial z^2}\right)+\sum_{i=1,2}-\frac{\hbar^2}{2m}\frac{\partial^2}{\partial x_i^2}\nonumber\\
        &+U_0 \cos^2\left(\frac{\pi x}{L}\right)+\frac{4\pi\hbar^2a_s}{m}\delta(\mathbf{r})\frac{\partial}{\partial r}(r\cdot),
        \label{eq; Hamiltonian}
    \end{align}
    where $\mathbf{r}_1$, $\mathbf{r}_2$ are the coordinates of two bosons, with $\mathbf{r}=\mathbf{r}_2-\mathbf{r}_1$, $U_0$ is the lattice depth, $L$ is the lattice constant,  and $a_s$ is the $s$-wave scattering length. As illustrated in Fig.~\ref{fig:Theo}(a), two bosons initially in $D$ band scatter into the final bands $(n_1, n_2)$ through the first-order scattering process, which depends on the overlapping integral defined as~\cite{PhysRevA.87.063638}:
    \begin{align}
        &\Gamma^{D,D}_{n_1,n_2}(q'_1,q'_2;q_1,q_2) \nonumber\\
        =&\int^L_0\mathrm{d}x\ u^*_{n_1,q_1}(x)u^*_{n_2,q_2}(x)u_{D,q'_1}(x)u_{D,q'_2}(x),
        \label{eq: overlap integral}
    \end{align}
    where $u_{n_i, q_i}$ is the periodic Bloch function for bosons in energy band $n_i$ with quasi-momentum $q_i$. Moreover, the scattering cross-section is given by:
    \begin{align}
        &\sigma(D,D;n_1,n_2)v=\frac{8\pi a_s^2L^2}{m}\int \mathrm{d}q\mathrm{d}q_1\mathrm{d}q_2\ \Xi(q_1,q_2)  \nonumber\\ &\theta(E_{D,q_1}+E_{D,q_2}-E_{n_1,(q_1+q_2)/2+q}-E_{n_2,(q_1+q_2)/2-q})\nonumber \\ & \times \left|\Gamma^{D,D}_{n_1,n_2}(q_1,q_2;(q_1+q_2)/2+q,(q_1+q_2)/2-q)\right|^2,
        \label{eq: cross-section}
    \end{align}
    where $v$ is the relative velocity of the two bosons, and $\Xi(q_1, q_2)$ is the quasi-momentum distribution of the two bosons, $\theta(x)$ is the Heaviside step function defined as $\theta(x) = 1$ when $x \geq 0$ and $\theta(x) = 0$ when $x < 0$. $E_{n_i, q_i}$ is the single-atom energy. This scattering cross-section quantifies the strength of the inter-band or intra-band scattering process, namely the scattering channel.

    For two bosons in $D$ band, the possible final bands after scattering include $(n_1,n_2)=(S,S),\ (P,P),\ (S,P),\ (P,D)$, and $(S,D)$. In the weakly interacting regime, where the momentum broadening effect is negligible, the scattering cross-sections can be calculated by approximating the two-body quasi-momentum distribution in the $D$ band as $\Xi(q_1, q_2) \approx \delta(q_1) \delta(q_2)$ in Eq.~\eqref{eq: cross-section}. Applying this approximation, the relative scattering cross-sections in the weakly interacting system with varying lattice depths are presented in Fig.~\ref{fig:Theo}(b). It shows that across all trap depths, the scattering cross-section for $(D,D) \rightarrow (P, P)$ (the dark blue solid line) is the largest, followed by $(D,D) \rightarrow (S, S)$ (the shallow blue solid line). Specifically, we demonstrate the relative cross-section when $U_0 = 10 E_r$ in Fig.~\ref{fig:Theo}(c), which shows the dominant scattering channel is $(D,D) \rightarrow (P, P)$ due to the large overlap between the wave functions~\cite{Shui:23}. It is noteworthy that $D$-band bosons undergoing the first scattering event may further experience secondary scattering, thus making the results of $D$-band scattering different from the theoretical predictions above.

    Given the $s$-wave scattering length $a_s$, the scattering rate of the $D$-band bosons, $K(D, D)$, can be calculated by summing over the cross-sections of all possible scattering channels:
    \begin{align}
        K(D, D)= n_D\sum_{n_1, n_2}\sigma(D, D; n_1, n_2) v = n_D R(U_0) a_s^2,
        \label{eq: K factor}
    \end{align}
    where $R(U_0)$ is a function of $U_0$ determined by the scattering cross-sections and $n_D$ is the density of $D$-band bosons held in the optical lattices. Therefore, the lifetime of $D$-band bosons, $\tau$, can be estimated by the initial scattering rate $K^i(d, d)$:
    \begin{align}
        \tau \propto \frac{1}{K^i(D, D)} = \frac{1}{n^i_D R(U_0)a_s^2},
        \label{eq: as^{-2}}
    \end{align}
    where $n^i_D$ is the initial density of bosons in the $D$ band after loading. 
    
    In this manner, we theoretically predict the factors influencing the scattering rates of $D$-band bosons and identify the dominant scattering channel. Experimental validation of these predictions requires low particle temperatures, rapid preparation of $D$-band occupation (relative to the timescale of scattering lifetimes), and the capability to directly control scattering lengths.

    \section{Experimental demonstration}\label{sec:experimentImplementation}
    
    To study the above collisional scattering phenomena, our experiments are performed with BECs of $^{6}\rm Li$ Feshbach molecules~\cite{doi:10.1126/science.1093280}, with each molecule constituted by two lithium atoms in the lowest hyperfine states $|F = 1/2,m_F=1/2\rangle\ $($|1\rangle $) and $|F = 1/2,m_F=-1/2\rangle\ $($|2\rangle $). The inter-molecule interaction strength can be set over a range by tuning the $s$-wave scattering length $a_{12}$ between atoms in states $|1\rangle$ and $|2\rangle$ via the Feshbach resonance~\cite{PhysRevLett.94.103201} and the $s$-wave scattering length between molecules is given by $a_s = 0.6a_{12}$~\cite{PhysRevLett.93.090404}.

    Figure.~\ref{fig:experimentalSetup}(a) shows the schematic of the experimental setup (full details provided in Ref.~\cite{PhysRevA.109.043313}). The mBECs of about $20\,000$ molecules are confined in the crossed optical dipole traps formed by a pair of far-red-detuned lasers in a vertical plane with a $30^{\circ}$ to each other. A pair of hollow electric coils produce the Feshbach Resonance magnetic field. The trapping frequencies are  $(\omega_x,\omega_y,\omega_z ) = 2\pi \times (39.5,187.8,195.0)$ Hz, where the $x$ axis refers to the horizontal direction where the plane trapping beams are located, $y$ the other horizontal direction, and $z$ the vertical direction. The one-dimensional optical lattice is formed by two counter-propagating beams of $\lambda = 1064\ \rm nm$ lasers resulting in a lattice constant $L = \lambda/2 = 532\ \rm nm$, with lattice potential $U(x) = U_0  \mathrm{cos}^2(\pi x/L)$ in the horizontal direction, where $U_0$ is the lattice depth. The characteristic lattice energy is $E_r = \hbar^2 \mathit{k_l}^2/2\mathit{m} $, where $k_l = \pi/L$ and $m$ is the mass of a $^{6}\rm Li_2$ molecule.

    The experimental time sequence is presented in Fig. \ref{fig:experimentalSetup}(b). After evaporative cooling, the degenerate gas is prepared at a temperature of $T/T_{\rm F}\simeq 0.1$. The Feshbach magnetic field is then adiabatically ($300 \ \mathrm{G/s}$) ramped to the target value and kept for an additional duration to stabilize. Then a nonadiabatic shortcut method is utilized to load particles from the harmonic trap into the $D$-band $\Gamma$ point ($q=0$) of the lattice. For different lattice depths, which are calibrated by Kaptiza-Dirac (KD) scattering~\cite{kapitza_dirac_1933}, a distinctive sequence of optical pulses with intervals is optimized to reach the target state with high fidelity~\cite{Zhou_2018,PhysRevResearch.6.023217}. For instance, a two-pulse sequence $(t_1^{\rm on},t_1^{\rm off},t_2^{\rm on},t_2^{\rm off} ) = (2.9,6.0,8.3,5.4)\ \rm \mu s$ is used for the lattice depth $U_0=10 E_r$ with the theoretical fidelity above $99.5\%$ in the non-interacting limit. This shortcut method can be applied to different lattice depths and interactions while maintaining high fidelity (see Appendix \ref{Sup:A} for details). After the loading process, the $D$-band mBECs stay in the lattice for a period of evolution time $t_{\rm evo}$. Then, we apply the band mapping method~\cite{PhysRevLett.87.160405} to read out band distribution by ramping down the lattice in the form of $e^{-t_{\rm map}/\tau}$, where $t_{\rm map}=100\ \mathrm{\mu s}$ and $\tau=50\ \mathrm{\mu s}$. This mapping process projects particles in different bands into the corresponding Brillouin zone (BZ). Thus, molecules at the $D$-band $\Gamma$ point are mapped to the boundary between the second and third BZs at $p=\pm 2\hbar k_l$. Finally, the optical dipole traps and the optical lattices are both turned off, and particles in different states expand during a TOF process, which is detected via standard absorption imaging.

    A typical distribution of $D$-band mBECs is shown in Fig.~\ref{fig:experimentalSetup}(c). This distribution is integrated along the y-axis to obtain a one-dimensional density distribution function. Here, we use a bimodal fitting method to obtain the number of molecules in the $D$-band, which is marked by the yellow dashed line in the bottom panel of Fig.~\ref{fig:experimentalSetup}(c). The function can be expressed as follows:
    \begin{align}
    	f(p) = A_0e^{-\frac{(p-p_0)^2}{2w_0^2}} + \sum^5_{i=1} A_i\left(1 - \frac{(p-p_i)^2}{w_i^2}\right)^2.
    	\label{eq: fit Li}
    \end{align}
    The first term represents the scattering halo, while the second term represents condensed particles at $p=0$ ($S$ band), $\pm \hbar k_l$ ($P$ band), and $\pm 2\hbar k_l$ ($D$ band). Here, $A_i$ are the amplitudes and $w_i$ are their widths. By integrating the corresponding terms, we determine the number of particles in the $D$ band ($N_D$) and the total particle number ($N$). The results for inter-molecule scattering length with $a_{s}= 655a_0$ (Fig.~\ref{fig:experimentalSetup}(c1)) and $1100a_0$ (Fig.~\ref{fig:experimentalSetup}(c2)) are displayed, where $a_0$ donates the Bohr radius ($0.0529$ nm). We observe that the larger interactions result in a more prominent scattering halo, which will be investigated in our later work. 

    Here, we define the strongly and weakly interacting regimes based on experiments. In the strongly interacting regime, where $a_s > 500 a_0$, there are additional effects beyond the excited-band scattering process, such as coherence loss and more pronounced halos during TOF expansion. Whereas in the weakly interacting regime, for $a_s < 100 a_0$, the halo and coherence loss are negligible. Despite these additional effects in the strongly interacting regime, the fitting process ensures that our measurements of the $D$-band proportion remain solid.

    The proportion of remaining condensed molecules in the $D$-band over time can be obtained by changing the evolution time $t_{\rm evo}$ in the lattices. As shown in Fig.~\ref{fig:experimentalSetup}(d), for $a_{s} = 655 a_0$ and $U_0=10E_r$, $\mathrm{P}_D = N_D / N$ normalized to the initial proportion $\mathrm{P}^i_D=0.7$ as a function of varied $\tau_{\rm evo}$ is shown by the blue dots, with $\tau_{\rm evo}$ changed every 0.1 $\rm ms$ up to $2\ \rm ms$. It is fitted by a red solid line to calculate the molecular $D$-band lifetime with $\mathrm{P}_D/\mathrm{P}^i_D=1/e$, and we get the $D$-band lifetime $\tau=(1.393 \pm 0.071)$ ms in this situation. In this way, we can measure the lifetimes and scattering rates of $D$-band molecules under different interactions and lattice depths.

    \section{Influence of interactions on excited-band lifetimes}\label{sec:interaction}

    According to Eq.~\ref{eq: as^{-2}}, $D$-band lifetime is primarily influenced by three factors: the $s$-wave scattering length $a_s$, the initial particle number density $n_{D}^{i}$, and $R(U_0)$, which is determined by the scattering cross-section as a function of $U_0$. The first two factors are directly related to interactions, while $U_0$ can also influence interactions in optical lattices to some extent. Therefore, in this section, we investigate how the $D$-band lifetime is affected by the interactions originating from the interplay between $a_s$ and $U_0$, and the resulting influence as reflected through its trend.

\begin{figure}
        \includegraphics[width=0.48\textwidth]{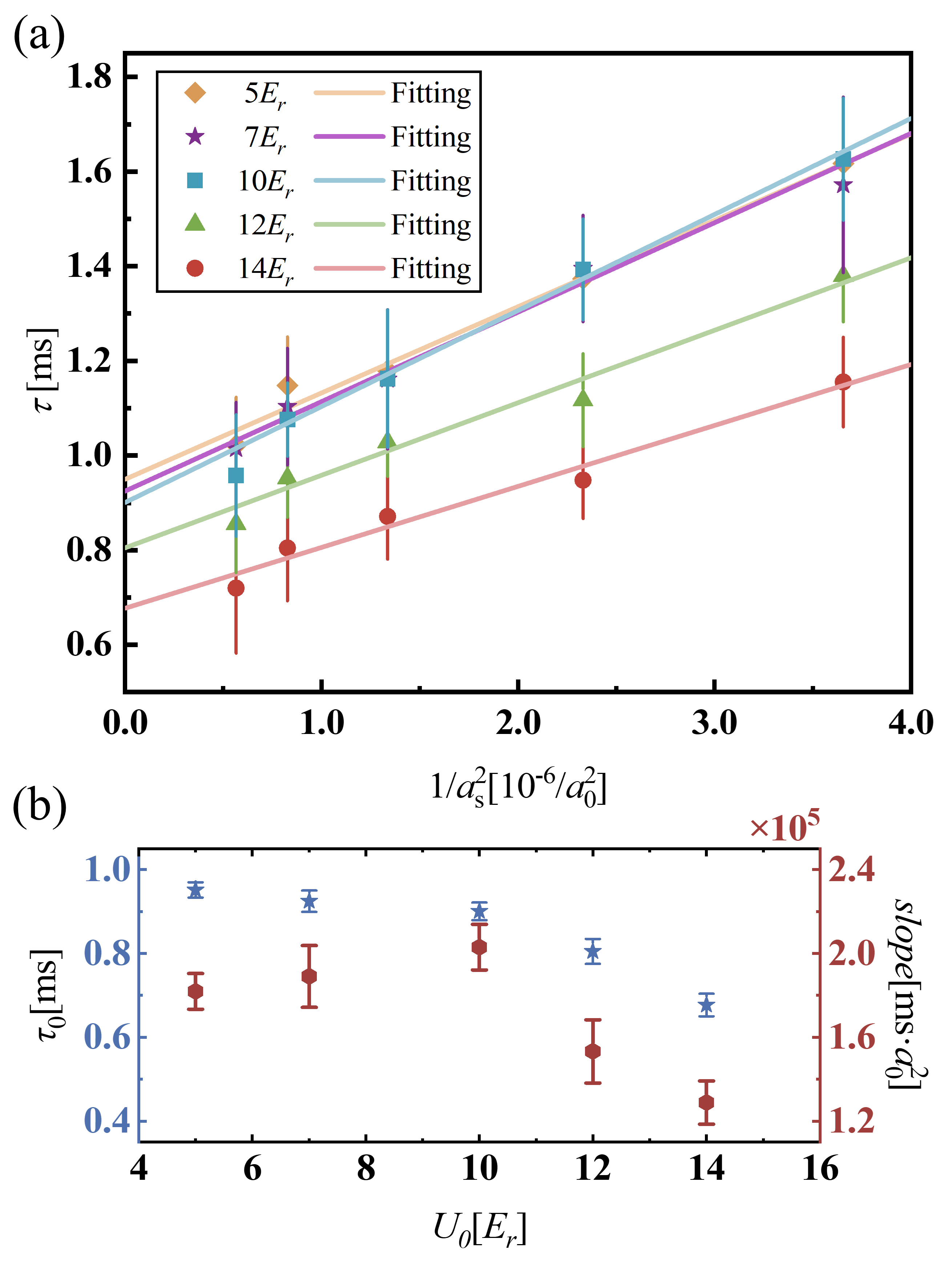}
        \caption{
            (a) $D$-band lifetimes under different interaction strengths ($n_D = 1.0 \times 10^{18} \ \rm m^{-3}$). The yellow diamonds, purple stars, blue squares, green triangles, and red circles indicate the results of $U_0=$ $5E_r$, $7E_r$, $10E_r$, $12E_r$, and $14E_r$. The corresponding lines are linear fitting of $\tau-1/a_{s}^{2}$. (b) The intercept of the $\tau-1/a_{s}^{2}$ fitting lines at different lattice depths are shown as $\tau_{0}$ by blue stars, representing the $D$-band lifetimes when interactions tend toward divergence. The red hexagons indicate the slopes of the $\tau-1/a_{s}^{2}$ fitting lines at different lattice depths.}
        \label{fig:interaction}
    \end{figure}   

    \subsection{The relationship between scattering length and \texorpdfstring{$D$}{}-band lifetime under varying lattice depths}
    To explore the relationship between the $D$-band lifetime and the $s$-wave scattering length, we measure lifetimes across different lattice depths with varying interaction strengths while maintaining the particle number density nearly constant ($n_D = 1.0 \times 10^{18} \ \rm m^{-3}$) by adiabatically tuning optical dipole traps. The experimental results are presented in the form of $\tau - 1/a_{s}^{2}$ to better assess the relationship, in Fig.~\ref{fig:interaction}(a). As an example with $U_0=10E_r$, we measured the $D$-band lifetime for $a_s=1330a_0$, $1100a_0$, $865a_0$, $655a_0$, and $523a_0$, which correspond to the five blue squares from left to right in Fig.~\ref{fig:interaction}(a), respectively. The error bars represent the fitting uncertainty of the lifetimes. By fitting, we can find a linear correlation between $\tau$ and $1/a_{s}^{2}$ across all lattice depths, with R-squares all above $0.98$. However, it is important to clarify that this relationship holds only in a certain interaction regime. In the non-interacting regime, the lifetime improves but does not approach infinity due to the non-uniformity of the external harmonic trap and collisions with the background gas. In the unitary regime, where the scattering length diverges, the energy band picture is no longer successful in this situation, leading the physics there to become even more complex.

    By analyzing the linear fitting parameters in Fig.~\ref{fig:interaction}(a), we observe an unexpected trend that exceeds the predictions of our theoretical model, as shown in Fig.~\ref{fig:interaction}(b). First, phenomenologically, the intercept of the fitting lines (termed $\tau_{0}$) represents the $D$-band lifetimes as interactions become even stronger while the energy band theory remains valid. $\tau_{0}$ (illustrated by the blue stars) decreases monotonically with increasing lattice depth $U_0$, indicating that deeper lattice depths shorten the $D$-band lifetime in this regime. This finding aligns with the intuitive physical image that a deeper lattice facilitates localization within a single lattice site and enhances interactions, thereby reducing the excited-band lifetimes. Second, the slopes reflect the rate of change of the scattering rate with the scattering length, which is essentially a function of $R(U_0)$. We observe a non-monotonic trend in the slopes as $U_0$ increases (illustrated by the red hexagons), with the slopes rising for $U_0 < 10E_r$ and decreasing for $U_0 > 10E_r$. It somewhat reflects the trend of $R(U_0)$ within the strongly interacting regime and will be explored in more detail in the subsequent study (Sec.~\ref{sec:optimal}).
    
    Moreover, we note that for stronger interactions ($a_s>1600 a_0$) with deeper lattices ($U_0>10E_r$), the experimental data do not yield reliable lifetime fittings due to the drastic decay in $D$-band occupation and influential scattering halos. These observations suggest that when $a_s \rightarrow +\infty$, $\tau$ gradually deviates from linearity and converges to zero, marking the failure of excited-band scattering theory. 
    
    To summarize, in the strongly interacting regime where the scattering length does not diverge, the relationship $\tau - \tau_{0} \propto 1/a_{s}^{2}$ holds consistently and can be supported by our experimental data, which means the scattering rate is proportional to $a_{s}^{2}$. This relation can be generalized to all excited energy bands with necessary corrections.

    \subsection{Variation of the \texorpdfstring{$D$}{}-band lifetime with lattice depth under different interactions}\label{sec:optimal}
    
\begin{figure}
        \includegraphics[width=0.48\textwidth]{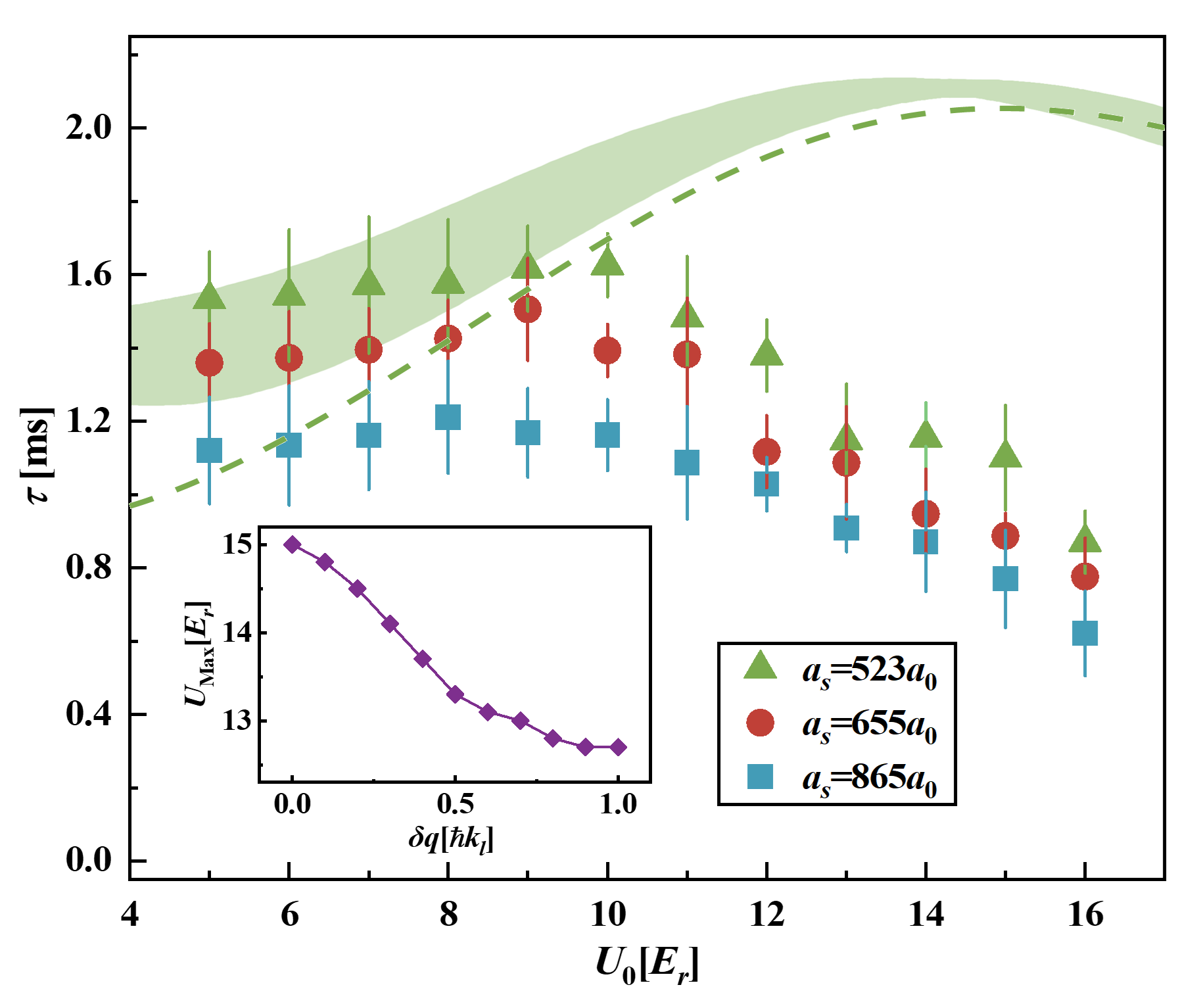}
        \caption{The $D$-band lifetimes with varying $U_0$ ($n_D = 1.0 \times 10^{18} \ \rm m^{-3}$) are shown under different values of $a_s$. The green triangles, red circles, and blue squares represent the experimental results for $a_s = 523a_0$, $655a_0$, and $865a_0$, respectively. The theoretical results with momentum broadening, $\delta q = 0$, and $0.1 \leq \delta q / \hbar k_l \leq 0.5$ are depicted by the green dashed line and the green shading, respectively. The subplot shows the theoretical peak lattice depth $U_{\rm Max}$ as a function of momentum width $\delta q$.}
        \label{fig:optimaldepth}
    \end{figure}

    In our previous work~\cite{Shui:23}, we predicted the existence of an optimal lattice depth for weakly interacting $D$-band bosons in various optical lattice configurations, including 1D lattices. However, its validity under stronger interaction regimes remains unclear. To explore this, we conduct experiments with varying $U_0$ at $a_s = 523a_0$, $655a_0$, and $865a_0$, as shown in Fig.~\ref{fig:optimaldepth}. The experimental results suggest that, for $U_0\le 10E_r$, the $D$-band lifetime remains largely unchanged with a slight increase, whereas for $U_0>10E_r$, it decreases monotonically. The lifetime is maximized at around $10E_r$, corresponding to the maximum slope in Fig.~\ref{fig:interaction}(b). Meanwhile, the experimental data shows that stronger interactions tend to shift the peak to a lower lattice depth, with the maximum lifetime at $10E_r$ for $a_s=523a_0$ shifting to $8E_r$ for $a_s=865a_0$. It suggests other interaction effects on the excited-band lifetime, leading to the absence of a universal optimal lattice depth for the $D$ band within the strongly interacting regime.

    By performing numerical calculations, we can obtain the lifetimes for different lattice depths at $a_s=523a_0$ (green dashed line in Fig.~\ref{fig:optimaldepth}). There is a significant discrepancy between the theoretical predictions and the experimental results. From a qualitative perspective, the discrepancy may be attributed to two primary factors. 
    
    The first factor is the non-negligible momentum broadening arising from the strong interaction strength, especially when $a_s>500a_0$. In this situation, the momentum distribution in the vicinity of the $\Gamma$ point can no longer be approximated as a $\delta$ function. We instead approximate the two-body quasi-momentum distribution as a Gaussian function:
    \begin{align}
        \Xi(q_1,q_2) = \frac{1}{2\pi \delta q^2} e^{-\frac{q_1^2+q_2^2}{2\delta q^2}},
        \label{eq: Gaussian}
    \end{align}
    where $q_i$ is the quasi-momentum of the two particles, and $\delta q$ is the quasi-momentum broadening width in the $D$ band. Using Eqs.~\ref{eq: cross-section} and \ref{eq: K factor}, we calculate the $D$-band lifetime as a function of lattice depth for different values of $\delta q$. The theoretical lifetime interval for $0.1 \leq \delta q / \hbar k_l \leq 0.5$ at $a_s=523a_0$, with the initial $D$-band density $n^i_D = 1.0 \times 10^{18} \rm m^{-3}$ (approximated by the BEC density), is shown in Fig.~\ref{fig:optimaldepth} by the green shading. As $\delta q$ increases, $\tau$ increases in the shallow lattice, and its maximum shifts to a lower $U_0$. This shifting behavior qualitatively agrees with our experimental results. However, it is important to note that $\delta q$ cannot be easily deduced from our experiment because the $D$-band particles separate into two clusters after band mapping, and the TOF process is affected by interaction effects. The observed wider distribution under stronger interactions may arise from the non-linear expansion due to the interaction effect rather than a genuinely larger $\delta q$, although higher $q$ states in the $D$ band can be occupied. 
    
    The second factor is the decoherence caused by interactions. It may explain the rapid reduction in lifetime when $U_0 > 10 E_r$. As $U_0$ increases, the localization of molecules at each lattice site induces the loss of superfluidity and reduces coherence, which may substantially increase the scattering rate of the excited-band state. Additionally, a larger $a_s$ tends to initiate this localization in a shallower lattice, contributing to the observed behavior of peak shift.

    In conclusion, these two factors can qualitatively account for the trend differences between the experimental results and theoretical predictions. However, a quantitative explanation requires further investigation.

\begin{figure*}
    \includegraphics[width=0.8\textwidth]{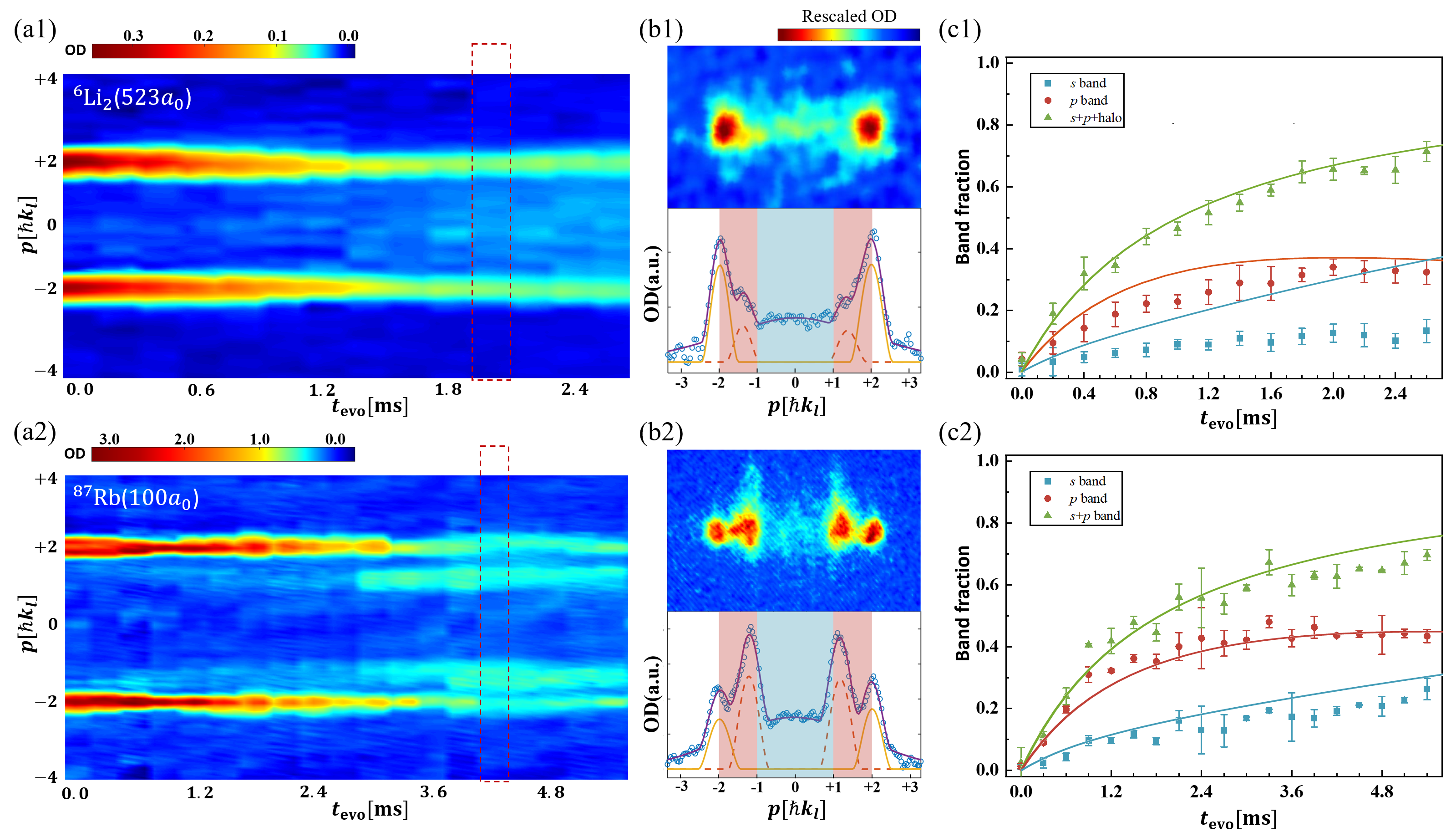}
    \caption{
		(a) The collisional scattering process of $D$-band molecules under $U_0=10E_r$. Band-mapping distributions are plotted against evolution time $t_{\rm evo}$. (b) (top) Raw images of the red dashed boxes in (a) and (bottom) the corresponding 1D density distribution along the lattice. The first and second BZs are indicated by blue and red shading, respectively. The $D$-band condensed parts are shown by yellow solid lines. The $P$-band condensed parts are marked by red dashed lines. (c) The variation curves of the particle numbers in different bands. Blue squares and red circles represent the scattered particle numbers in the $S$ and $P$ bands, respectively. Green triangles represent all scattered particles except for those in the $D$ band. The solid lines represent the solution of the rate equation accounting for the secondary scattering. (a1)(b1)(c1) $^{6}\rm Li_2$ molecules ($a_s = 523a_0$, $n_D=1.0\times10^{18}\ \rm m^{-3}$, $t_{\rm map}=0.1\ \rm ms$). (a2)(b2)(c2) $^{87}\rm Rb$ atoms ($a_s = 100a_0$, $n_D=1.0\times10^{20}\ \rm m^{-3}$, $t_{\rm map}=0.5\ \rm ms$).}
	\label{fig:duration}
\end{figure*}

 \section{Discussion on the Collisional scattering process in 1D optical lattices}\label{sec:process}
    \subsection{Scattering process in a strongly interacting system}
    To distinguish the scattering channel of $D$-band molecules in a 1D optical lattice with remarkable interactions, we perform the $D$-band experiment with a molecular number density $n_d=1.0\times10^{18}\ \rm m^{-3}$ and a lattice depth of $U_0=10E_r$ to investigate the collisional scattering process in the $D$ band. To minimize the scattering halo resulting from in-trap and time-of-flight collision, we choose the minimum scattering length $a_s = 523 a_0$. The band-mapping distributions against $t_{\rm evo}$ are shown in Fig.~\ref{fig:duration}(a1).
    
    Despite the minimum scattering length utilized in our experiment, a scattering halo still covers the lower band distribution after band mapping. A typical TOF image is presented in Fig.~\ref{fig:duration}(b1) with $t_{\rm evo}=2 \mathrm{ms}$. Aside from the $D$-band parts (the yellow solid line), we observe two additional condensates (the red dashed line) in the second BZ through the fitting. Note that particles in the $P$ band possess a negative effective mass, causing them to gradually shift to $\pm \hbar k_l$ in the presence of an external harmonic trap. Thus, these condensates correspond to $P$-band particles. Since the scattered particles consist of both condensates and thermal clouds, we determine band fractions by integrating the first term in Eq.~\ref{eq: fit Li} within the first BZ ($S$ band, the blue shading) and second BZ ($P$ band, the red shading) as well as the $S$-band parts (hardly any) and $P$-band parts (the red dashed line) in the second term, respectively. Moreover, the proportion of scattering halos at $t_{\rm evo}=0$ ms is subtracted from the total particle number to eliminate disturbance from the scattering halos.
    
    As shown in Fig.~\ref{fig:duration}(c1), it is evident that the increase of particle number in the $P$ band is significantly higher than in the $S$ band, suggesting that the dominant scattering channel is $(D,D) \rightarrow (P,P)$. It exhibits a consistent conclusion with the prediction in Sec.~\ref{sec:theoretic}. However, interactions disrupt the information of scattered particles, thus the conclusion regarding the substantial accumulation in the $P$ band can only be inferred through fitting rather than observed directly.

    \subsection{Scattering process in a weakly interacting system}
    To observe scattering channels and the secondary scattering process more clearly, we repeat the experiment in a weakly interacting ${}^{87}\mathrm{Rb}$ system ($a_s=100a_0$) due to the difficulty of further reducing interactions in the $^{6}\mathrm{Li}_2$ system. The experimental results, with the atomic number density $n_D=1.0\times10^{20}\ \rm m^{-3}$ and lattice depth $U_0=10E_r$, are shown in Fig.~\ref{fig:duration}(a2). A more detailed description of this system is given in Ref.~\cite{PhysRevA.104.L060601}, and the scattering halo effect is mostly suppressed here. After an evolution time of $t_{\rm evo} = 1\ \text{ms}$, a significant distribution of particles at $\pm \hbar k_l$ states is directly observed, as demonstrated by Fig.~\ref{fig:duration} for $t_{\rm evo}=4.2\ \text{ms}$. We present the obtained band fractions in Fig.~\ref{fig:duration}(c2) and it also suggests that the $(D, D) \rightarrow (P, P)$ process is the relatively dominant scattering channel. 
    
    Additionally, we observe a slight decline of $P$-band distribution after 3 ms, which becomes more pronounced after 5.4 ms. It may be attributed to the enhancement of secondary scattering. To explain this, we develop a rate equation (see Appendix \ref{Sup:B} for details) by considering all scattering channels from the $D$ band and the secondary scattering from the $P$ band, to simulate the variation curves of band fractions in the $S$, $P$, and $D$ bands. The solutions of the rate equation are represented by the solid lines in corresponding colors in Fig.~\ref{fig:duration}(c2). The solutions fit well with the experimental data, especially of the proportion of $P$-band scattered particles, suggesting that the excited-band scattering theory is quite robust in a weakly interacting system. Nonetheless, the theoretical model does not work well in our $^{6}\rm Li_2$ system (solid lines in Fig.~\ref{fig:duration}(c1)), which means it needs further modification to become more applicable for systems in the strongly interacting regime.

    \section{Conclusion}\label{sec:Conclusion}

    In this work, we explore the effects of interactions on the collisional scattering of $D$-band $^{6}\mathrm{Li}_2$ molecules in a 1D optical lattice with the shortcut loading method. By measuring the lifetimes of $D$-band particles under various interaction strengths, we experimentally investigate the squared relationship between the scattering rate of excited-band particles and the $s$-wave scattering length by measuring $D$-band lifetimes under different interactions. Meanwhile, interactions also affect the trend of $D$-band lifetime varying with lattice depth by increasing momentum broadening and reducing coherence. The former leads to a decrease in the lattice depth at the peak point, while the latter causes a sharp decline to lifetime in relatively deep lattices ($U_0>10E_r$). Consequently, there is no universal optimal lattice depth in the strongly interacting regime.
    
    Moreover, we analyze and discuss the scattering process of $D$-band particles. In the strongly interacting $^{6}\mathrm{Li}_2$ system, we observe the dominant scattering channel $(D,D) \rightarrow (P,P)$ by fitting, but scattering halos induced by interactions hinder our direct detection of detailed scattering phenomena. For comparison, we conduct the same experiments in a weakly interacting $^{87}\mathrm{Rb}$ system and observe more details. The rate equation developed in this study accurately calculates the evolution of band fractions when accounting for secondary scattering. 
    
    This work enhances our understanding of the interplay between interactions and collisional scattering of excited-band particles in optical lattices, paving the way for future research into many-body physics in lattice systems. 

    \section*{Acknowledgements}
	This work is supported by the National Natural Science Foundation of China (Grants No. 92365208, No.11934002, and No. 11920101004), National Key Research and Development Program of China (Grants No. 2021YFA0718300 and No. 2021YFA1400900). C.L. is supported by the Austrian Science Fund (FWF) through the ESPRIT grant 'Entangled Atom Pair Quantum Processor' [grant DOI: 10.55776/ESP310 (EAPQuP)].The authors would like to thank Chi Zhang for the discussion related to coding.

    \appendix
    \addcontentsline{toc}{section}{Appendices}\markboth{APPENDICES}{}
    \begin{subappendices}

    \section{\texorpdfstring{$D$}{}-band Shortcut Sequences}\label{Sup:A}
    
    The shortcut method is a robust way to load BEC from the harmonic trap into optical lattices. In our previous work \cite{Zhou_2018,PhysRevLett.121.265301,PhysRevA.87.063638}, we have used the method to load bosons into $S$ band and excited bands of 1D, 2D or 3D lattice. The basic principle of shortcut is that the evolution operators $L_{\vec{k}}(t)$ of momentum states $\left |\vec{k} \right \rangle$ are different between the lattice on and off. As shown in Fig.\ref{fig:experimentalSetup} (c), after several laser pulses, the final state is:
		\begin{eqnarray}
		\label{state_aftertwopulse}
		\left | \psi_f \right \rangle = \sum_{\vec{k}} \prod \limits_{i=1}^n L^{\rm off}_{\vec{k}}(t^{\rm off}_{i}) L^{\rm on}_{\vec{k}}(t^{\rm on}_i) \times \left |\vec{k} \right \rangle,
		\end{eqnarray}
    where $n$ is the number of pulses, $L^{\rm on/off}_{\vec{k}}(t)$ is the evolution operator when the lattice is on/off.
    \begin{table}[htp]
    \caption{The shortcut sequences to load bosons into the \texorpdfstring{$D$}{} band in 1D optical lattices with different lattice depth.}
    \label{tab:initial_seq}
    \begin{tabular}{cccccccccc}
    \hline
    \specialrule{0em}{1pt}{1pt}
        {$V_0$}    & {$t^{\rm on}_1[\mu \rm s]$}    &{$\ \ t^{\rm off}_1\ \ $}    &{$\ \ t^{\rm on}_2\ \ $}    &{$\ \ t^{\rm off}_2\ \ $}      & {Theoretical fidelity}  \\
    \hline
    \specialrule{0em}{1pt}{1pt}
    {$5 E_r$} & {$4.4$}    & {$11.2$}    & {$7.3$}    & {$5.7$}      & {$99.992\%$} \\
    {$7 E_r$} & {$9.1$}    & {$2.3$}    & {$7.6$}    & {$23.7$}     & {$99.995\%$} \\
    {$10 E_r$} & {$2.9$}    & {$6.0$}    & {$8.3$}    & {$5.4$}    & {$99.992\%$} \\
    {$12 E_r$} & {$9.9$}    & {$14.2$}    & {$0.5$}    & {$13.7$}    & {$99.988\%$} \\
    {$14 E_r$} & {$5.0$}    & {$10.3$}    & {$9.7$}    & {$22.4$}    & {$99.860\%$} \\
    \hline
    \end{tabular}	
    \end{table}
		
    By choosing the number of pulses and pulse length, we can optimize the final state $\left | \psi_f \right \rangle$ to aimed state $\left | \psi_a \right \rangle$. The fidelity is defined by $|\left \langle \psi_f|\psi_a \right \rangle |^2$ to describe the loading efficiency. In the experiment, the optimized sequence has two pulses to load atoms into the $D$ band of 1D optical lattices, and the pulse sequences of different lattice depths are shown in Table \ref{tab:initial_seq}. The theoretical fidelity is calculated under non-interaction conditions.
		
    If interactions are considered, the loading fidelity of the shortcut method decreases somewhat. As an example, the theoretical fidelity at $U_0=10E_r$ with strong interactions is calculated by the Gross-Pitaevskii equation (GPE) and demonstrated in Table \ref{tab:inter_seq}. Although there is a drop, we can not observe condensates in other bands experimentally. Thus we conclude that this method is still valid within the strongly interacting regime.
    \begin{table}[htp]
    \caption{The \texorpdfstring{$D$}{} band fidelity of shortcut method in 1D optical lattices with different interaction strengths ($U_0=10E_r$).}
    \label{tab:inter_seq}
    \begin{tabular}{c|cccccc}
    \hline
    \specialrule{0em}{1pt}{1pt}
        {$a_s$}    & {$523a_0$}    &{$655a_0$}    &{$865a_0$}    &{$1100a_0$}      & {$1330a_0$}      & {$1600a_0$}\\
       \hline 
        {Fidelity}    & {$99.992\%$}    &{$99.992\%$}    &{$99.991\%$}    &{$99.991\%$}      & {$99.990\%$}      & {$99.988\%$}\\
    \hline
    \end{tabular}	
    \end{table}

    \section{Rate equation of the excited-band scattering process}\label{Sup:B}
    
    In this section, we develop a rate equation to evaluate the scattering process in the excited band. For the first scattering event, particles in the $D$ band scatter into the $S$ and $P$ bands with different quasi-momenta. We define the first scattering rate per unit density as follows:
    \begin{align}
        &R_{1,DP} = \sigma(D,D; P,P)v + \frac{1}{2}\sigma(D,D; D,P)v, \nonumber\\
        &R_{1,DS} = \sigma(D,D; S,S)v,
        \label{eq: first scattering}
    \end{align}
    where $\sigma(n_1,n_2; n'_1,n'_2)$ is defined in Eq.~\ref{eq: cross-section}, and the channels $(D,D)\rightarrow(S,D)$, $(S,P)$ are neglected here due to their minimal scattering cross-sections.

    After the first scattering event, $P$-band particles can continue to scatter into the $S$ band, a process we call secondary scattering. We now evaluate the secondary scattering rate, $R_{2, PS} = \sigma(P,P; S,S)v$, from the $P$ band at different $q$ states to the $S$ band. The $P$-band quasi-momentum distribution, $f_P(q)$, after first scattering from the $D$ band can be approximated by the overlap integral:    
    \begin{align}
        f_P(q) = \frac{\left|\Gamma^{D,D}_{P,P}(0,0; q,-q)\right|^2}{\int \mathrm{d}q \left|\Gamma^{D,D}_{P,P}(0,0; q,-q)\right|^2},
    \end{align}
    which is the normalized differential scattering cross-section of the $D$ band to the $P$ band. Note that the $(D,D)\rightarrow(D,P)$ process is neglected in the secondary scattering process for simplicity. With this distribution, we set $\Xi(q_1,q_2) = f_P(q_1)f_P(q_2)$ in $\sigma(P,P; S,S)v$ and obtain $R_{2,PS}$. Combining both the first and secondary scattering processes, the differential equations governing the evolution of the scattering process are given by:
    \begin{align}
        &\frac{\mathrm{d}n_D}{\mathrm{d} t} = -R_{1,DP}n_D^2 - R_{1,DS}n_D^2,\nonumber\\
        &\frac{\mathrm{d}n_P}{\mathrm{d} t} = R_{1,DP}n_D^2 - R_{2,PS}n_P^2,\nonumber\\
        &\frac{\mathrm{d}n_S}{\mathrm{d} t} = R_{1,DS}n_D^2 + R_{2,PS}n_P^2,
        \label{eq: RE}
    \end{align}
    with the initial condition: $n_D(0)=n^i_D,\ n_S(0)=n_P(0)=0$. The solution of these equations, with and without $R_{2, PS}$, is displayed in Fig.~\ref{fig:duration}(b). The experimental data suggest that secondary scattering plays a significant role in practice.

    \end{subappendices}

%References
    \bibliographystyle{apsrev4-2}
    \bibliography{ref}

%\begin{refcontext}[sorting = none]
%\printbibliography
%\end{refcontext}
 
\end{document}